\documentclass[aps,prd,floatfixy,superscriptaddress,showpacs,showkeys]{revtex4}[18pt]
\usepackage{amssymb}
\usepackage{amsmath}
\usepackage{amsfonts}
\usepackage{epsfig}
\usepackage{graphicx}
\usepackage{graphics}
\usepackage{subfigure}
\usepackage{color}
\usepackage{verbatim}
\def\Lam{\Lambda_c}

\def\Lam+{\Lambda_c^+}
\def\Lamp+{\Lambda_c^{'+}}
\def\blam+{{\bar{\Lambda}_c^+}}

\newcommand{\eq}{\begin{eqnarray}}
\newcommand{\en}{\end{eqnarray}}

\usepackage[normalem]{ulem}

\renewcommand\sout{\bgroup \color{red} \ULdepth=-.5ex \ULset}

\def\ds{d^*(2380)}
\def\eq{\begin{eqnarray}}
\def\en{\end{eqnarray}}
\def\ds{d^*}
\def\d12{D_{12}}

\begin{document}

\title{\huge\bf  On the form factors of $d^*(2380)$}
\author{Yubing Dong}
\affiliation{Institute of High Energy Physics, Chinese Academy of
Sciences, Beijing 100049, China}
\affiliation{Theoretical Physics
Center for Science Facilities (TPCSF), CAS, Beijing 100049, China}
\affiliation{School of Physical Sciences, University of Chinese
Academy of Sciences, Beijing 101408, China}
\author{Pengnian Shen}
\affiliation{College of Physics and Technology, Guangxi Normal
University, Guilin  541004, China}
\affiliation{Institute of High
Energy Physics, Chinese Academy of Sciences, Beijing 100049, China}
\affiliation{Theoretical Physics Center for Science Facilities
(TPCSF), CAS, Beijing 100049, China}
\author{Zongye Zhang}
\affiliation{Institute of High Energy Physics, Chinese Academy of
Sciences, Beijing 100049, China} \affiliation{Theoretical Physics
Center for Science Facilities (TPCSF), CAS, Beijing 100049, China}
\affiliation{School of Physical Sciences, University of Chinese
Academy of Sciences, Beijing 101408, China}
\iffalse
\author{M.~Bashkanov}
\affiliation{School of Physics and Astronomy, University of
Edinburgh, Peter Guthrie Tait Road, Edinburgh EH9 3FD, United
Kingdom}
\fi
\date{\today}
\begin{abstract}
In order to explore the possible physical quantities for judging different structures of the newly observed
resonance $d^*(2380)$, we study its electromagnetic form factors. In addition to the electric charge monopole
$C0$, we calculate its electric quadrupole $E2$, magnetic dipole $M1$, and six-pole $M3$ form factors
on the base of the realistic coupled $\Delta\Delta+CC$ channel $\ds$ wave function with both the $S$- and
$D$-partial waves. The results show that the magnetic dipole moment and electric quadrupole deformation of $\ds$
are 7.602 and $2.53\times 10^{-2}~\rm{fm}^2$, respectively. The calculated magnetic dipole moment in the naive
constituent quark model is also compared with the result of $D_{12}\pi$ picture. By comparing with partial results
where the $\ds$ state is considered with a single $\Delta\Delta$ and with a $D_{12}\pi$ structures, we find that
in addition to the charge distribution of $\ds$, the magnetic dipole moment and magnetic radius can be used to
discriminate different structures of $\ds$. Moreover, a quite small electric quadrupole deformation indicates that
$\ds$ is more inclined to an slightly oblate shape due to our compact hexaquark dominated structure of $\ds$ .
\end{abstract}
\maketitle
\section{Introduction}
\noindent\par

Since the dibaryon state was proposed more than 50 years ago, the existence of the dibaryon has become one of the
hot topics in particle and nuclear physics. Among those states, $H$ particle and $\ds$ states were involved most.
In particular, the $d^*$ state has intensively been studied by various models from the hadronic degrees of
freedom to the quark degrees of freedom more than half century. Its mass prediction was ranged from a few MeV
to several hundred MeV. Searching such a state has also been considered as one of the aims in several
experimental projects. However, no confidential results were released until 2009. Since then, a series of
experimental studies for $\ds$ was carried out in the study of ABC effect by CELSIUS/WASA and WASA@COSY
Collaborations.~\cite{CELSIUS-WASA, Adlarson:2011bh, Adlarson:2012fe,Adlarson:2014pxj}. Various double-pion and
single-pion decays of $\ds$, including invariant mass spectra, Dalitz plots, Argon plots, in the $pn$ and
$pA$ reactions, the analyzing power of the neutron-proton scattering and etc., have been carefully measured
and analyzed. It was found that the results cannot be explained by the contribution either from the intermediate
Roper excitation or from the t-channel $\Delta\Delta$ contribution, except introducing an intermediate new
resonance. Then, the discovery of a new resonance with a mass of about $2370\sim 2380~\rm{MeV}$, a width of
about $70\sim 80~\rm{MeV}$, and the quantum numbers of $I(J^P)=0(3^+)$ was announced~\cite{CELSIUS-WASA,
Adlarson:2011bh, Adlarson:2012fe, Adlarson:2014pxj}. Since the baryon number of the resonance is 2, it is
believed that such a state is just the $d^*$ state which has been hunted for several decades, and, in general,
can be explained by either "an exotic compact particle" or "a hadronic molecule state". \\

It should be emphasized that the threshold (or cusp) effect may not be so significant in the $\ds$ case as that
in the XYZ particle case due to the fact that the observed mass of $\ds$ is about $80~\rm{MeV}$ below the
$\Delta\Delta$ threshold and about $70~\rm{MeV}$ above the $\Delta\pi N$ threshold~\cite{Chen:2016qju,
Guo:2017jvc,Dong:2017gaw}. In addition, if  $d^*$ does exist, it contains at least 6 light quarks, and
it is also different from the XYZ particles which contain heavy flavor.\\

Following the reports of Refs. ~\cite{CELSIUS-WASA, Adlarson:2011bh, Adlarson:2012fe,Adlarson:2014pxj},
many theoretical models for the structure of $d^*$ have been developed or proposed. Up to now, there are
mainly two structural schemes which attract considerable attention of physicists. One of them assumes
that the $d^*$ state has a compact structure, and may be an exotic hexaquark dominated state whose mass
is about $2380-2414~\rm{MeV}$ and width about $71~\rm{MeV}$, respectively
~\cite{Yuan,Brodsky, Huang:2014kja,Huang:2015nja,Dong,Dong1,Dong2,Dong2017}. The other
one, in order to explain the upper limit of the single-pion decay width of $\ds$~\cite{Clement2017}, proposes that
the $d^*$ state is basically a molecular-like hadronic state with a $\alpha~[\Delta\Delta]+(1-\alpha)~[D_{12}\pi]$
mixing structure  ($\alpha=5/7$) ~\cite{Gal1}, which originates from a
three-body $\Delta N \pi$ resonance assumption, where
the pole position of the resonance locates around ($2363\pm 20$)~$+{\it i} (65\pm 17)$ MeV,
~\cite{Gal:2013dca,Gal:2014zia} and a $D_{12}\pi$ molecular-like model, where the mass and width of the resonance
are pre-fixed to be $2370~\rm{MeV}$ and $70~\rm{MeV}$, respectively~\cite{Kukulin}. Although the experimental data
can be explained by using either scheme, the described structures of $\ds$ are quite different. Therefore, it is
necessary to seek other physical observables which would have distinct values for different interpretations  so
that with the corresponding experimental data one would be able to justify which one is more reasonable.\\

It is well-known that with the help of the electromagnetic probe, electromagnetic form factors become indispensable
physical quantities in revealing the internal structure of a complicated system. For example, the electromagnetic
form factors of a nucleon provide us the charge and magnetic distributions inside the nucleon. This fact exhibits
the structure of the nucleon where a three quark core is surround by the pion cloud. It also tells that the
charge and magnetic radii of the nucleon can be extracted by the slops of the charge and magnetic distributions
of the nucleon at $Q=0$ (where Q stands for the momentum transfer). The accurately measured charge radius of the
proton does justify the structure of the nucleon. Furthermore, in a spin-1 system, for instance a deuteron or a
vector $\rho$-meson, the charge, magnetic and quadrupole form factors can also reveal its intrinsic structures,
such as its charge and magnetic distributions and the quadrupole deformation. Consequently, the electromagnetic
form factors might also be discriminating quantities for studying the inner structure of the higher spin particle.
In particular, for the $\ds$ state, if there is a considerably large hidden color component (HCC) in it, we found
that, although such a component does not contribute to its hadronic strong decay in the leading-order
calculation, but it can play a rather important role in the charge distribution calculation~\cite{Dong, Dong1,
Dong2,Dong:2017mio}. The resultant charge distribution of $d^*$ with a compact 6-quark structure is quite
different from that having a $D_{12}\pi$ (or $\Delta\pi N$) structure~\cite{Dong:2017mio}. Therefore, we believe
that the charge distribution of $\ds$ can serve as one of the criteria for judging its structure.\\

In general, a spin-3 particle has $2S+1=7$ electromagnetic form factors, $C0$ (charge monopole form factor),
$C2$ (or $E2$, electric quadrupole form factor), $C4$, and $C8$ for electric form factors and $M1$, $M3$, and $M5$
for magnetic form factors. Therefore, in order to understand the structure of $\ds$, the spin-3 particle, except
the charge distribution (namely, the charge monopole form factor $C0$) calculated in our previous
paper~\cite{Dong:2017mio}, we are going to study the other lower rank form factors of $\ds$, such as its electric
quadrupole  $E2$, magnetic dipole $M1$, and magnetic six-pole $M3$ form factors, with a compact $\Delta\Delta+CC$
coupled-channel structure on the base of our chiral SU(3) constituent quark model.\\

This paper is organized as follows. In Sect. II, the wave function of $\ds$ in the chiral SU(3) constituent
quark model is briefly introduced. Sec. III is devoted to the electromagnetic form factors and the multipole
decomposition of the electromagnetic current of the $\ds$ resonance. Our numerical results and a short
summary will be presented in Sec. IV.\\

\section{Wave functions of $d^*(2380)$ in chiral constituent quark model}
\par\noindent\par

In studying possible dibaryons in 1999, a $\Delta\Delta+CC$ structure of the $\ds$ state with $(I(J^P))=(0(3^+))$,
where $I$, $J$, $P$ are isospin, spin, and parity, respectively, was firstly proposed ~\cite{Yuan}. With the
assumption of such a structure, its mass of $2384~\rm{MeV}$ was predicted~\cite{Yuan}. In recent years, a series
of sophisticated studies on the structure and decay properties of $\ds$ has further been performed, and a compact
picture for it, an exotic hexaquark dominated state, was deduced~\cite{Yuan, Huang:2014kja,
Huang:2015nja,Dong,Dong1,Dong2,Dong:2017mio}. In order to get a meaningful result for a 6-quark system, those
calculations were dynamically carried out on the quark degrees of freedom by using a chiral SU(3) constituent quark
model. In this strong interaction model, the effective quark-quark interaction induced by the exchange of the
chiral fields receives the contributions from the pseudoscalar, scalar, and vector chiral fields, respectively.
The model parameters are determined in such a way that the stability condition and the properties of nucleon, the
mass splitting between the nucleon and $\Delta$, the spectra of low-lying baryons, the static properties of
deuteron, and the phase shifts of the nucleon-nucleon scattering can be ensured. With these pre-fixed model
parameters, we believe that the model has considerable prediction power~\cite{Yu:1995ag,Zhang:1997ny}.\\

\begin{figure}[htbp]
\begin{center}
\includegraphics[width=10cm,height=8cm] {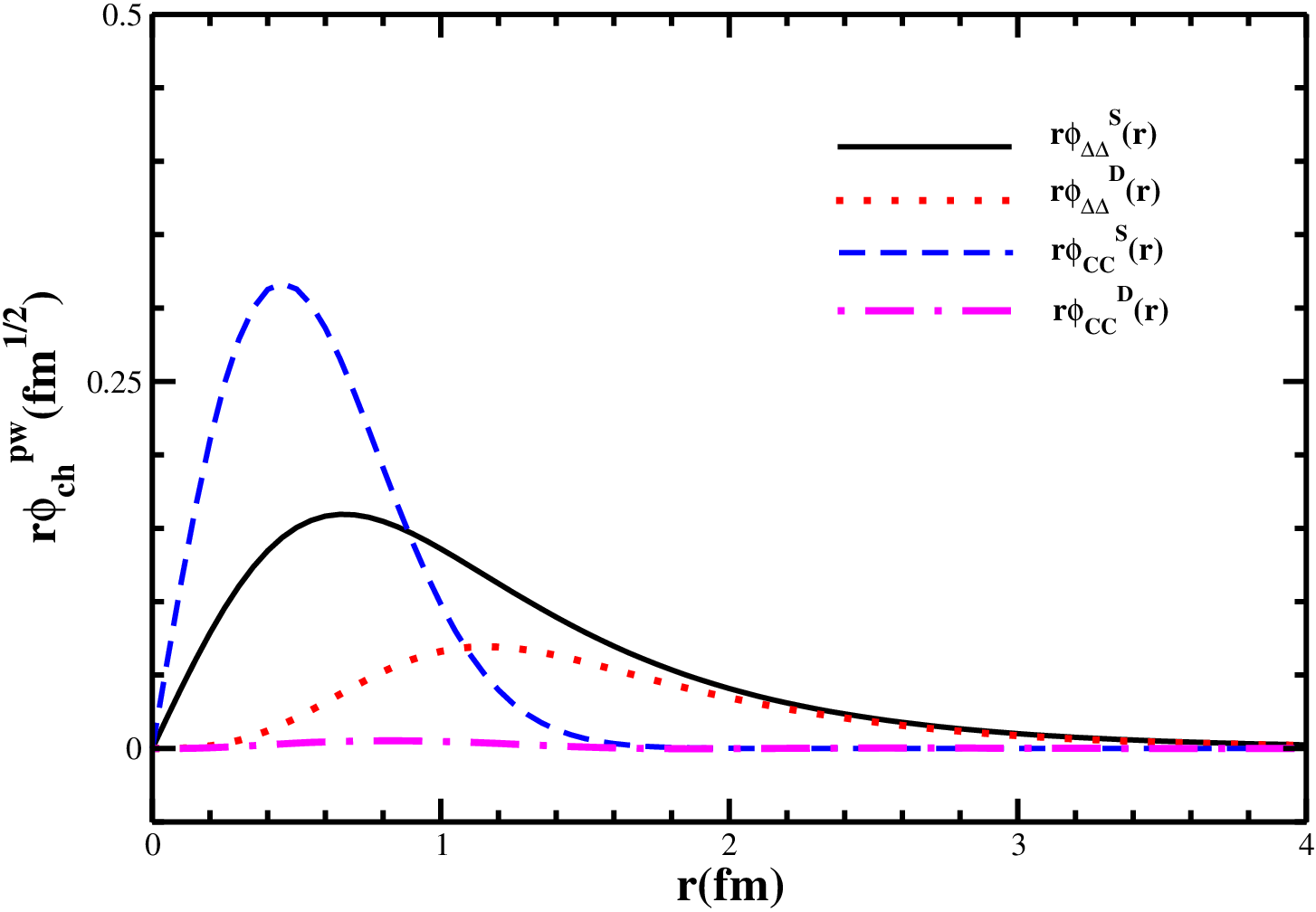}
\caption{Channel wave functions in the $\ds$ system. The black solid, red dotted, blue dashed, and pink
dotted-dashed curves describe the contributions from the $S$- and $D$-waves in the $\Delta\Delta$ channel, and the
$S$- and $D$-waves in the $CC$ channel, respectively. }
\label{Fig1}
\end{center}
\end{figure}

In the practical calculation for $\ds$, we use the well-established Resonating Group Method (RGM) which has
frequently been applied to the studies of nuclear physics and hadronic physics, especially where the clustering
phenomenon exists. In the RGM framework, if we assume again that the $\ds$ state has a $\Delta\Delta+CC$ structure,
the full 6-quark wave function reads
\begin{eqnarray}
\label{eq:wf} \Psi_{6q}={\cal A}\Big
[\phi_{\Delta}(\vec{\xi}_1,\vec{\xi}_2)\phi_{\Delta}(\vec{\xi}_4,\vec{\xi}_5)\eta_{\Delta\Delta}(\vec{r})
+\phi_{C}(\vec{\xi}_1,\vec{\xi}_2)\phi_{C}(\vec{\xi}_4,\vec{\xi}_5)\eta_{CC}(\vec{r})\Big
]_{S=3, I=0}^{C=(00)},
\end{eqnarray}
where ${\cal A}=1-9P_{36}$ is the antisymmetrizer in the orbital (O), spin (S), isospin (F), and color (C) spaces,
respectively, due to the Pauli exclusion principle, $\phi_{\Delta(C)}$ denotes the internal wave function of the
$\Delta(C)$ cluster with $\xi_i(i=1,2~(4,5))$ being the internal Jacobi coordinates in the first (second) cluster,
and $\eta_{\Delta(C)}$ stands for the relative wave function between the two $\Delta(C)$ clusters, which
will be determined by solving the RGM equation. However, due to the effect of the quark exchange, two components in
eq.~(\ref{eq:wf}) are not orthogonal to each other. To see the role of each component in the properties of
the $\ds$ state, an orthogonalization procedure for these two components should be taken. It can be done by making
projections
\begin{eqnarray}
\label{eq:wf1}
\chi_{\Delta\Delta}(\vec{r})&=&<\phi_{\Delta}(\vec{\xi}_1,\vec{\xi}_2)\phi_{\Delta}(\vec{\xi}_4,\vec{\xi}_5)\mid
\Psi_{6q}>\\ \nonumber
\chi_{CC}(\vec{r})&=&<\phi_{C}(\vec{\xi}_1,\vec{\xi}_2)\phi_{C}(\vec{\xi}_4,\vec{\xi}_5)\mid
\Psi_{6q}>,
\end{eqnarray}
respectively. Clearly, the newly achieved functions, called channel wave functions, are orthogonal to each other
and contain all necessary quark-exchange effects. Then, the wave function of the system could be expressed as
\begin{eqnarray}
\label{eq:wf2} |\ds(S_{\ds}=3,M_{\ds})>&=&~~~\big
[|\Delta\Delta>_{S_{\ds}=3,M_{\ds}}\chi_{\Delta\Delta}^{S,0}\big
]_{S_{\ds}=3,M_{\ds}}
+\big [|\Delta\Delta>_{S_{\ds}=3,M_{S}}\chi_{\Delta\Delta}^{D,m}\big ]_{S_{\ds}=3,M_{\ds}} \nonumber \\
&&~+\big[|CC>_{S_{\ds}=3,M_{\ds}}\chi_{CC}^{S,0}\big
]_{S_{\ds}=3,M_{\ds}}+\big [|CC>_{S_{\ds}=3,M_S}\chi_{CC}^{D,m}\big
]_{S_{\ds}=3,M_{\ds}}\\ \nonumber &=&\sum_{ch=\Delta\Delta,
CC}~~~\sum_{pw=S,D} \Big
[|ch>_{S_{\ds}=3,M_S}\chi_{ch}^{pw,m_l}(\vec{r})\Big
]_{S_{\ds}=3,M_{\ds}}
\end{eqnarray}
with $ch=\Delta\Delta$ and $CC$ denoting the constituents of the component, $M_{\ds}$ representing the magnetic
quantum number of spin $S_{\ds}$, $pw=l=0$ and $2$ representing the $S$ and $D$ partial waves ($pw$) between
the two clusters, respectively, and $m_l$ being its magnetic quantum number. Again these four channel wave
functions are orthogonal to each other. In comparison with our previous calculations for the strong decay and
charge distribution where the contribution from the $D$-wave is ignored because it is negligibly small, here we
include the relative $D$-wave in the calculations of the higher multipole form factors, such as $E2$, and $M3$
since those values are closely related to the matrix elements of the high-rank operators.  The
relative wave functions in eq.~(\ref{eq:wf2}), with $\chi_{ch}^{S,0}(\vec{r}) = \phi_{ch}^S(\mid
r\mid)Y_{00}(\Omega_r)$ and $\chi_{ch}^{D,m}(\vec{r})=\phi_{ch}^D(\mid r\mid )Y_{2m}(\Omega_r)$, are displayed in
Fig.~\ref{Fig1}, respectively.
\vspace{1cm}
The probabilities of $S$- and $D$-waves in the $\Delta\Delta$ and $CC$ channels are determined by
\begin{eqnarray}
\label{eq:prob}
{\cal P}_{ch}^{pw,m}=\int d^3r \big |\chi_{ch}^{pw,m}(\vec{r})\big
|^2,
\end{eqnarray}
and their magnitudes are shown in Tab.~\ref{Tab:prob}.
\begin{table}[htbp]
\caption{Probabilities of various components of $\ds$}
\begin{center}
\begin{tabular}{|c||c|c|c|c|} \hline
Channel       &\multicolumn{2}{|c|} {$\Delta\Delta$}
              &\multicolumn{2}{|c|}{$CC$} \\ \hline
partial wave  &$S$ &$D$  &$S$  &$D$\\ \hline ~~~probability ${\cal
P}^{pw.}_{ch}(\%)$ ~~~&~~~31.19~~~ &~~~0.50~~~ &~~~68.31~~~ &~~~$\sim$ 0.002~~~\\
\hline
\end{tabular}
\label{Tab:prob}
\end{center}
\end{table}
From Fig.~\ref{Fig1} and Tab.~\ref{Tab:prob}, one sees that comparing with corresponding $D$-waves, the
$S$-wave in both $\Delta\Delta$ and $CC$ channels are overwhelmingly dominant, and the $D$-waves are negligibly
small. The probability of the $S$-wave in the $CC$ channel is about 2 times larger than that in the
$\Delta\Delta$ channel which is essential for our understanding of the partial widths in the double-pion and
single-pion decays of $\ds$, and consequently of a narrow total width in our assumption of the compact structure of
$\ds$~\cite{Huang:2014kja, Huang:2015nja,Dong,Dong1,Dong2,Dong2017,Dong:2017mio}.\\

\section{Multi-pole decomposition and electromagnetic form factors of $d^*$}
\noindent\par

$d^*(2380)$ is a spin-3 particle, it has $2s+1=7$ form factors. In general, a traceless rank-3 tensor,
$\epsilon_{\alpha\beta\gamma}$, can be employed to describe the spin-3 field. Clearly,
$\epsilon_{\alpha\alpha\beta}=0$, $\epsilon_{\alpha\beta\gamma}=\epsilon_{\beta\alpha\gamma}$, and
$p^{\alpha}\epsilon_{\alpha\beta\gamma}=0$. \\

In the one-photon exchange approximation, the general form of the electromagnetic current of the  $3^+$
particle can be written as~\cite{Dong:2017mio}
\begin{eqnarray}
\label{eq:current} {\cal J}^{\mu}=\big (\epsilon^*\big
)^{\alpha'\beta'\gamma'}(p'){\cal
M}^{\mu}_{\alpha'\beta'\gamma',\alpha\beta\gamma}\epsilon^{\alpha\beta\gamma}(p)
\end{eqnarray}
with the matrix element
\begin{eqnarray}
\label{eq:matrix} {\cal
M}^{\mu}_{\alpha'\beta'\gamma',\alpha\beta\gamma}&=& \Bigg
[G_1(Q^2){\cal P}^{\mu}\Big [g_{\alpha'\alpha}\Big
(g_{\beta'\beta}g_{\gamma'\gamma}+g_{\beta'\gamma}g_{\gamma'\beta}\Big
)+permutations\Big ]\\ \nonumber &&+G_2(Q^2){\cal P}^{\mu}\Big
[q_{\alpha'}q_{\alpha}\big [g_{\beta'\beta}g_{\gamma'\gamma}+
g_{\beta'\gamma}g_{\gamma'\beta}\big ]+permutations\Big ]/(2M^2)\\
\nonumber &&+G_3(Q^2){\cal P}^{\mu}\Big
[q_{\alpha'}q_{\alpha}q_{\beta'}q_{\beta}g_{\gamma'\gamma}+permutations\Big
]/(4M^4)
\\ \nonumber
&&+G_4(Q^2){\cal
P}^{\mu}q_{\alpha'}q_{\alpha}q_{\beta'}q_{\beta}q_{\gamma'}q_{\gamma}/(8M^6)\\
\nonumber &&+G_5(Q^2)\Big [ \Big
(g_{\alpha'}^{\mu}q_{\alpha}-g_{\alpha}^{\mu}q_{\alpha'}\Big)
\Big(g_{\beta'\beta}g_{\gamma'\gamma}+g_{\beta'\gamma}g_{\beta'\gamma}\Big
)+permutations \Big ]\\ \nonumber &&+G_6(Q^2)\Big [ \Big
(g_{\alpha'}^{\mu}q_{\alpha}-g_{\alpha}^{\mu}q_{\alpha'}\Big )
\Big(q_{\beta'}q_{\beta}g_{\gamma'\gamma}+q_{\gamma'}q_{\gamma}g_{\beta'\beta}
+q_{\beta'}q_{\gamma}g_{\gamma'\beta}+q_{\gamma'}q_{\beta}g_{\gamma\beta'}\Big
)\\ \nonumber &&~~~~~~~~~~~~~~+permutations\Big ]/(2M^2)\\ \nonumber
&&+G_7(Q^2)\Big [\Big
(g_{\alpha'}^{\mu}q_{\alpha}-g_{\alpha}^{\mu}q_{\alpha'}\Big
)q_{\beta'}q_{\beta}q_{\gamma'}q_{\gamma}+permutations\Big ]/(4M^4)
\Bigg ],
\end{eqnarray}
where $M$ is the mass of $d^*(2380)$, ${\cal P}=p'+p$ (with $p'$ and $p$ being the momenta of the outgoing and
incoming $\ds$, respectively), and $G_{i}(Q^2),~i=1,2,\dots,7$, are the seven electromagnetic form factors
which depend on the momentum transfer square $Q^2=|\vec{q}|^2$. The gauge invariant condition
\begin{eqnarray}
\label{eq:gauge} q_{\mu}{\cal
M}^{\mu}_{\alpha'\beta'\gamma',\alpha\beta\gamma}=0
\end{eqnarray}
should also be fulfilled, as well as the time-reversal invariance. In general, the physical form factors,
such as the charge monopole $C0$, electric quadrupole $E2$, octupole $C4$, and twelve-pole $C6$ form factors,
as well as the magnetic dipole $M1$, six-pole $M3$ and ten-pole $M5$ form factors, can be constructed
by the combinations of the seven electromagnetic form factors $G_{i}(Q^2),~i=1,2,...7$.\\

The multipole decompositions of the electromagnetic currents, as well as the electromagnetic form factors of a
particle with spin-2 or with arbitrary spin, have been explicitly discussed in
Refs.~\cite{Spehler:1991yw,Aliev:2010ac, Lorce:2009bs, Lorce:2009br}. According to those analyses,
in the quark degrees of freedom, the time component of the photon-$d^*$ electromagnetic current, in the
instantaneous approximation, is $J^0=\sum_{i=1}^6 j_i^0$ with $j_i^0$ denoting the time component of the
photon-quark electromagnetic current for the $i$-th quark. The electric charge $l$-th multipole form factor
of $\ds$ reads
\begin{eqnarray}
\label{eq:GE}
G^E_l(Q^2)=\frac{(2M_{\ds})^l}{e}\sqrt{\frac{4\pi}{2l+1}}\frac{(2l+1)!!}{l!Q^l}{\cal
I}_{El}(Q^2),
\end{eqnarray}
with $e$ being the unit of charge and
\begin{eqnarray}
\label{eq:Inte} {\cal I}_{El}(Q^2)&=&<\ds\mid \sum_{i=1}^6\int
d^3r\big [d^3X\big ]e_ij_l\Big (Q\mid\vec{r}_i-\vec{R}\mid\Big
)Y_{l0}(\Omega_{r_i})\mid\ds>\\ \nonumber &=&3<\ds\mid \int d^3r\big
[d^3X\big ]\Big [e_3j_l\Big (Q\mid\vec{r}_3-\vec{R}\mid\Big )
Y_{l0}(\Omega_{\vec{r}_3-\vec{R}})\\ \nonumber
&&~~~~~~~~~~~~~~~~~~~~~~~~~+e_6j_l\Big (Q\mid\vec{r}_6-\vec{R}\mid\Big
)Y_{l0}(\Omega_{\vec{r}_6-\vec{R}})\Big ]\mid\ds>,
\end{eqnarray}
where $\big [d^3X\big ]=d^3\rho_1d^3\rho_2d^3\lambda_1d^3\lambda_2$, $\rho_1$, $\rho_2$, $\lambda_1$, and
$\lambda_2$ are the conventional Jacobi variables in the two clusters, and $j_l$ represents the $l$-th
spherical Bessel function. \\

The multipole decomposition of the space component of the electromagnetic current in the momentum space gives~\cite{Lorce:2009bs, Lorce:2009br}
\begin{eqnarray}
\label{eq:mtsp}
<d^*|\rho^M(\vec{q})|d^*>=e\sum_{l=0}^{+\infty}i^l\tau^{l/2}
\frac{l+1}{{\tilde C}_{2l-1}^{l-1}}G_{Ml}(Q^2)Y_{l0}(\Omega_q),
%~~~~\tau=\frac{Q^2}{4M^2},
\end{eqnarray}
where $\rho^M(\vec{q})$ denotes the magnetic density of the system  with $\tau=\frac{Q^2}{4M^2}$, and
\begin{eqnarray}
 \tilde{C}^k_n=\left\{ \begin{array}{ll}
\frac{n!!}{k!!~(n-k)!!},  &~~~~~~~~
n\geq k \geq -1, \\
0, &~~~~~~~~otherwise. \end{array}  \right.
\label{eq:perm}
\end{eqnarray}
If we only consider the quark-photon coupling, we can write the magnetic density as
$\rho^M(\vec{r})=\sum_{i=1}^6\vec{\nabla}\cdot \big (\vec{j}_i(r)\times\vec{r}_i\big )$
with $\vec{j}_i(r)$ and $\vec{r}_i$ being the current and position vectors for the
$i$-th quark in the coordinate space, and $\rho^M(\vec{q})=\sum_{i=1}^6\vec{\nabla}\cdot \big [\big
(e_i\vec{\sigma}_i\times\vec{q}\big )\times\vec{q}\big ]=2\sum_{i=1}^6e_i\vec{\sigma}_i\cdot\vec{q}$ with
$\vec{\sigma}_i$, $e_i$, and $\vec{q}$ being the Pauli matrix, the charge for the $i$-th quark and the
transferred momentum, respectively. Then, we have
\begin{eqnarray}
\label{eq:mtmag} <d^*|\rho^M(\vec{q})|d^*>&=&\sum_{i=1}^6<\ds\mid
\rho_i^M(\vec{q})\mid\ds>
\\ \nonumber
&=&\frac{6i}{2m_q}<\ds\mid\big [ e_3\vec{\sigma}_3\cdot\vec{q}
+e_6\vec{\sigma}_6\cdot\vec{q}\big ]\mid\ds>.
\end{eqnarray}
By assuming that the form factors are the functions of the momentum transfer square  $Q^2$ in the one-photon
exchange approximation, we have~\cite{Lorce:2009bs, Lorce:2009br}
\begin{eqnarray}
\label{eq:GM1} G_{M1}(Q^2)=-\int
d\Omega_qY^*_{10}(\Omega_q)\frac12\frac{i}{e\sqrt{\tau(Q)}}
\sqrt{\frac{3}{4\pi}}<d^*|\rho^{M}(\vec{q})|d^*>,
\end{eqnarray}
for $M1$ and
\begin{eqnarray}
\label{eq:GM3} G_{M3}(Q^2)=+\int
d\Omega_qY^*_{30}(\Omega_q)\frac{i}{2e\tau(Q)\sqrt{\tau(Q)}}
\sqrt{\frac{7}{4\pi}}\frac54<d^*|\rho^{M}(\vec{q})|d^*>,
\end{eqnarray}
for $M3$, respectively. \\

\section{Numerical results and discussions}

By using the wave functions of $\ds$ in eq.~(\ref{eq:wf2}), the charge monopole $C0$, electric quadrupole $C2$,
magnetic dipole $M1$, and magnetic six-pole $M3$ form factors for $\ds$ are calculated. It should be mentioned
that since the charge monopole form factor (or the charge distribution) of $\ds$ has already been discussed
explicitly in our previous paper~\cite{Dong:2017mio}, we do not reiterate the relevant result in detail in this
work. It is shown that the charge distribution receives the dominant contribution from the $S$ partial wave,
namely the $S-S$ matrix elements of the $\Delta\Delta$ and $CC$ components. Based on the fact that the slop of
the charge distribution is  related to the root-mean-square radius ($rms$) of the $\ds$ system, it is
found that compared to the $D_{12}\pi$ (or $\Delta\pi N$) structure, as well as to a single $\Delta\Delta$
structure, a compact hexaquark dominated structure for $\ds$, which is deduced from our coupled $\Delta\Delta+CC$ channel RGM calculation, has a much smaller $rms$~\cite{Dong:2017mio}. \\

\vspace{.5cm}
\begin{figure}[htbp]
\begin{center}
\includegraphics[width=9cm,height=7cm] {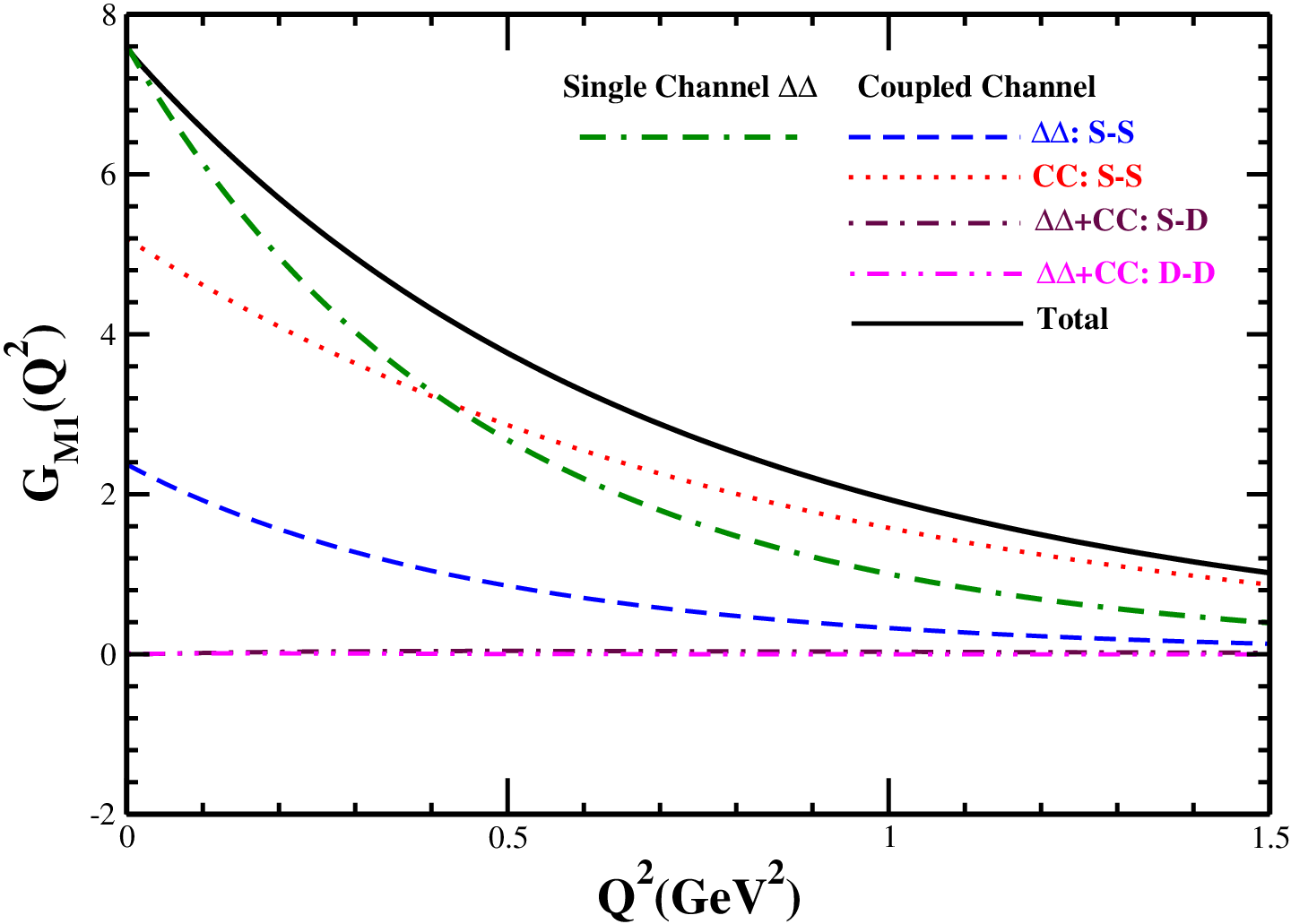}
\end{center}
\caption{\label{Fig2}The magnetic dipole form factors $M1$ of
$\ds$.}
\end{figure}

The magnetic dipole form factor $G_{M1}(Q^2)$ of $\ds$ is plotted in Fig.~\ref{Fig2}.  This form factor
respectively receives the contributions from the $S-S$  matrix elements of the $\Delta\Delta$ and $CC$
components, which are described by the blue-dashed and red-dotted curves in Fig.~\ref{Fig2}. Other
contributions from the $S-D$ matrix element (the off-diagonal matrix element between the $S$-wave and $D$-wave
functions) and the $D-D$ matrix element (the diagonal matrix element between $D$-wave functions) are negligibly
small compared to the $S-S$ components. These features can also be corroborated by the purple-dotted-dashed and
pink-double-dotted-dashed curves in Fig.~\ref{Fig2}. Clearly, the major contribution cames from the $S$-wave
of the $CC$ component, however the contribution from the $S$-wave of the $\Delta\Delta$ component is also sizable.
This is because that the probability of the $CC$ component is almost twice of that of the $\Delta\Delta$
component.\\

For a comparison, the calculated magnetic dipole form factor of $\ds$ with a
single channel $\Delta\Delta$ structure is demonstrated by the green
double-dashed-dotted curve also in Fig.~\ref{Fig2}. Furthermore, the
magnetic dipole momentum of $\ds$, $\mu_{\ds}$, can be extracted
from the magnetic dipole form factor at zero momentum transfer
$G_{M1}(Q^2=0)$. The obtained magnetic dipole moment of $\ds$
in the coupled channel $\Delta\Delta+CC$ cases is about 7.602 in
unit of $e$. Comparing with the proton and neutron magnetic dipole
moments of 2.79 and -1.91, respectively, this value is
understandable, because the number of quarks in $\ds$ is twice of
that in the proton or neutron. Moreover, the calculated magnetic dipole
moment of $\ds$ with a single $\Delta\Delta$ structure is about
7.612 which is almost the same as that of the $\ds$ state with a
$\Delta\Delta+CC$ structure. The tiny difference between two
magnetic dipole moments with different structures may be due to the
different amount of $D$-wave contributions. In addition, it should
be particularly stressed that the contribution from the off-diagonal
matrix element between the $\Delta\Delta$ and $CC$ channels vanishes
since the former has two colorless clusters and the latter has two
colored clusters, and the electromagnetic interaction is
color-independent.\\

In the naive constituent quark model (NCQM),  it is known that the magnetic moments of the proton and
neutron are about
$M_N/m_q\sim 3$ and $-2M_N/3m_q\sim -2$, respectively. These values roughly agree with the experimental data of
$2.79$ and $-1.91$. In the $\ds$ case, we find that, in the naive quark model, the contributions from the
$\Delta\Delta$ and $CC$ components are all proportional to $M_{\ds}/m_q$ due to $I(J^P)=0(3^+)$ for $\ds$ and
the quantum numbers of the $\Delta$ and $C$ clusters. Consequently, the magnetic moment of $\ds$, which relates to its magnetic form factor at the real photon limit, is
\begin{eqnarray}
\label{eq:M1} G_{M1}^{NCQM}(0)=\Big [{\cal P}^S_{\Delta\Delta}+{\cal
P}^S_{CC}\Big ]\frac{3M_{\ds}}{M_N}\sim \frac{3M_{\ds}}{M_N}=7.62,
\end{eqnarray}
where we approximately take $\Big [{\cal P}^S_{\Delta\Delta}+ {\cal P}^S_{CC}\Big ]\sim 1$  for the $S$-wave
as shown in Tab.~\ref{Tab:prob}, and $m_q\sim M_N/3$. This magnetic moment is very close to the
calculated value of 7.602 obtained from $G_{M1}(0)$ mentioned above. From the results for the proton, neutron
and $\ds$, one may believe that the magnetic moment of a particle estimated in the naive constituent quark model
can be taken as a qualitative reference in the study of the hadronic magnetic moment. The absolute ratio of
the calculated magnetic moment of the $\Delta\Delta$ component to that of the $CC$ component in our approach is
\begin{eqnarray}
\label{eq:ratio}
R^{\Delta\Delta}_{CC}=\frac{G_{M1}^{\Delta\Delta}}
{G_{M1}^{CC}}=\frac{2.37}{5.20}=0.4558,
\end{eqnarray}
which is almost the same as the probability ratio of the two component ${\cal P}_{\Delta\Delta}^{S}/ {\cal
P}_{CC}^{S}=31.19\%/68.31\%=0.4566$ shown in Tab.~\ref{Tab:prob}. Moreover, in the naive constituent quark model,
by using the same method, we obtain the magnetic moment of the $\ds$ state with a single $\Delta\Delta$ structure as
\begin{eqnarray}
\label{eq:M1DelDel}
{\tilde \mu}_{\ds}^{\Delta\Delta}~=~{\tilde G}_{M1}^{\Delta\Delta}(0)~\sim
\frac{3M_{\ds}}{M_N}=7.62.
\end{eqnarray}
This value is the same as that with a compact hexaquark dominated structure, which is understandable because the
averaged magnetic moment in the $\Delta\Delta$ component is the same as that in the $CC$ component. \\

In addition, the magnetic moment of $\ds$ with a $D_{12}\pi$ interpretation can also be calculated in this way.
We know that the spin of pion is zero, the contribution from the orbital angular moment between the $D_{12}$ and
$\pi$ systems vanishes. Then, the magnetic moment of $d^*$ comes from the $D_{12}$ cluster only. Therefore,
the obtained magnetic moment of $\ds$ in the naive constituent quark model is
\begin{eqnarray}
\label{eq:M1D12pi}
\mu_{\ds}^{D_{12}\pi}~=~G_{M1,\ds}^{D_{12}\pi}(0)~\sim
\frac{2M_{\ds}}{M_N}=5.07.
\end{eqnarray}
From the above obtained values for the different inner structures of $\ds$, one sees that the magnetic moment can
also serve as a quantity to distinguish between the compact hexaquark dominated structure (or the $\Delta\Delta$
structure) and the $D_{12}\pi$ structure, but not between the $\Delta\Delta+CC$ compact hexaquark dominated
structure and the $\Delta\Delta$ structure.\\

Furthermore, we know that the slop of $G_{M1}$ is  related to the magnetic radius of $\ds$. To see the
different contributions to the magnetic radius from the $\Delta\Delta$ component and the $CC$ component, we
check the slops of $G_{M1}^{\Delta\Delta}(Q^2)$ and $G_{M1}^{CC}(Q^2)$. They are
\begin{eqnarray}
\label{eq:magrad} -\frac{d}{dQ^2}G_{M1}^{\Delta\Delta}(Q^2)\Big
|_{Q^2\to 0}&=&5.014~\rm{GeV}^{-2}=0.195~\rm{fm}^2,\\ \nonumber
-\frac{d}{dQ^2}G_{M1}^{CC}(Q^2)\Big |_{Q^2\to
0}&=&6.139~\rm{GeV}^{-2}=0.239~\rm{fm}^2,
\end{eqnarray}
and their ratio is
\begin{eqnarray}
\label{eq:ratiomag}
R=\frac{\frac{d}{dQ^2}G_{M1}^{\Delta\Delta}(Q^2)\Big |_{Q^2\to
0}}{\frac{d}{dQ^2}G_{M1}^{CC}(Q^2)\Big |_{Q^2\to 0}}\sim 0.816.
\end{eqnarray}
This ratio contains the contributions from ${\cal
P}^S_{\Delta\Delta}$ and ${\cal P}_{CC}^S$, as well as from the
$Q^2$ dependent wave functions of the $\Delta\Delta$ and $CC$
components. The obtained value is remarkably different with the
probability ratio $R^{\Delta\Delta}_{CC}~\sim~0.4558$, which reveals
a fact that although the probability of the $\Delta\Delta$ component
is much smaller than that of the $CC$ component, but the normalized
magnetic radius of the $\Delta\Delta$ component is larger than that
of the $CC$ component, namely comparing with the $CC$ component, the
wave function of the $\Delta\Delta$ component distributes in a wider
range. As a final result, when $\ds$ has a compact hexaquark
dominant structure, the obtained slop of the magnetic form
factor $G_{M1}$ at the zero momentum transfer is
\begin{eqnarray} \label{eq:slopmag}
-\frac{d}{dQ^2}G_{M1}(Q^2)\Big |_{Q^2\to 0}&=&0.434~\rm{fm}^2,
\end{eqnarray}
and corresponding magnetic radius is
\begin{eqnarray}
\label{eq:magradius} \Bigg[\Big |-\frac{d}{dQ^2}G_{M1}(Q^2)\Big
|_{Q^2\to 0}\Big |\Bigg]^{1/2}&=&0.659~\rm{fm}.
\end{eqnarray}
In addition, in the single channel $\Delta\Delta$ case, we have the
magnetic radius being
\begin{eqnarray}
\label{eq:magrad1}
\Bigg [ \Big |-\frac{d}{dQ^2}G^{S}_{M1}(Q^2)\Big
|_{Q^2\to 0}\Bigg ]^{1/2}=0.896~\rm{fm}.
\end{eqnarray}
Clearly, one sees that the magnetic radius of $\ds$ with a single $\Delta\Delta$ structure is apparently larger
than that with a compact hexaquark dominated structure. Therefore, we believe that the magnetic radius of $\ds$
can serve as a physical quantity to distinguish between the $\Delta\Delta+CC$ and $\Delta\Delta$ structures of
$\ds$. Moreover, we stress that the magnetic feature of $\ds$ is consistent to the phenomenon revealed
in the case of the charge distribution of $\ds$, and the charge radius of $\ds$ ~\cite{Dong:2017mio} is
slightly larger than its magnetic radius. These characters also appeared in the experimental measurements for deuteron as has been discussed in Ref.~\cite{Afanasev:1998hu}.

\par\noindent\par
\vspace{1.5cm}
\begin{figure}[htbp]
\begin{center}
\includegraphics[width=9cm,height=7cm] {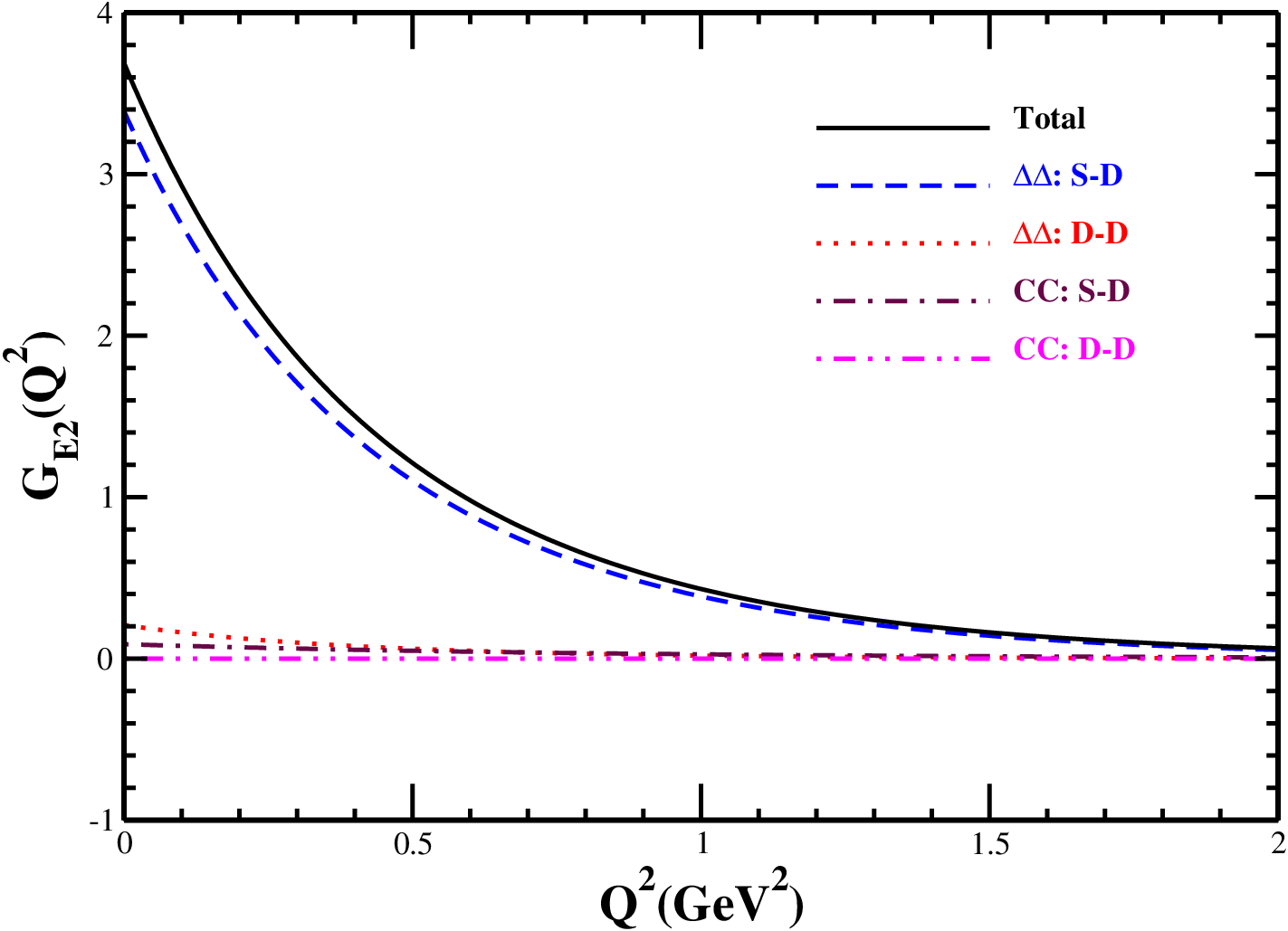}
\end{center}
\caption{\label{Fig3}The quadrupole form factors of  $\ds$.}
\end{figure}

%\vspace{0.5cm}
Our calculated electric quadrupole form factor $G_{E2}$ is shown in Fig.~\ref{Fig3}. In this figure, the
blue-dashed and red-dotted, purple-dotted-dashed, and pink-double-dotted-dashed curves describe the
contributions from the matrix elements between the $S$- and $D$-waves and the $D$- and $D$-waves of the
$\Delta\Delta$ component and from the matrix elements between the $S$- and $D$-waves and the $D$- and
$D$-waves of the $CC$ component, respectively. It should be mentioned that in this rank-2 operator case,
the diagonal matrix element between $S$-waves ($S-S$) does not contribute. The dominant contribution to
$G_{E2}$ comes from the off-diagonal matrix element between the $S$- and $D$-waves ($S-D$ or $D-S$) of the
$\Delta\Delta$ component. Since the probability of the $D$-wave of the $CC$ component is much smaller than
that of the $\Delta\Delta$ component as shown in table I, the contribution from the $D$-wave of the $CC$
component is negligibly small. Moreover, we have $G_{E2}(0)=\frac{M_{\ds}^2}{e}~Q_{20}^{\ds}$, where
$Q_{20}^{\ds}$ denotes the quadrupole deformation of $\ds$. Our calculation shows that such a deformation
is about $Q_{20}^{\ds}=2.53\times 10^{-2}~\rm{fm}^2$, which is much smaller than that of the deuteron
$Q_{20}^d=0.259~\rm{fm}^2$. This is because that the dominant contribution for the electric quadrupole
moment comes from the $S-D$ matrix element of the colorless-cluster component, namely the $\Delta\Delta$
component in $\ds$, as well as from that of the $p-n$ component in the deuteron case, however, the
probability of the $\Delta\Delta$ component in $\ds$ is only about 1/3 and the probability of the $p-n$
component in deuteron is almost 1, meanwhile the probability of the $D$-wave in the $\Delta\Delta$
component of $\ds$ (about 0.5\%) is much smaller than that in the $p-n$ component of deuteron (about 5\%).
These quadrupole deformations also indicate that $\ds$ is more inclined to a slightly oblate shape.
Therefore, our $\ds$ looks a more compact and spherical-shape due to its wave function.\\

It is known that the deformations of the nucleon and $\Delta$ give the $E2/M1$ ratio for the $\gamma N\to \Delta$
transition amplitude, which is one of the significant observables for judging different models. Here, one can
also consider the deformation in the $\Delta$ wave function. According to the previous analyses (see for example
Refs. ~\cite{Capstick:1992uc, Capstick:1992xn,Shen:1997}), the mixing coefficient for the component
$\Delta ^4D_s(\frac{3}{2})^+$ in the $\Delta$ resonance is about $-0.11$, the probability of such a configuration
is about 1.2\%, and the obtained $E2/M1 \simeq -1.0$\% for the $\Delta N$ transition. We can check the effect of
the deformation of $\Delta$ on the quadrupole moment. Our numerical calculation shows that this effect provides
a suppression of about 0.25\% to the quadrupole moment of $\ds$.\\

\vspace{0.5cm}
\begin{figure}[htbp]
\begin{center}
\includegraphics[width=9cm,height=7cm] {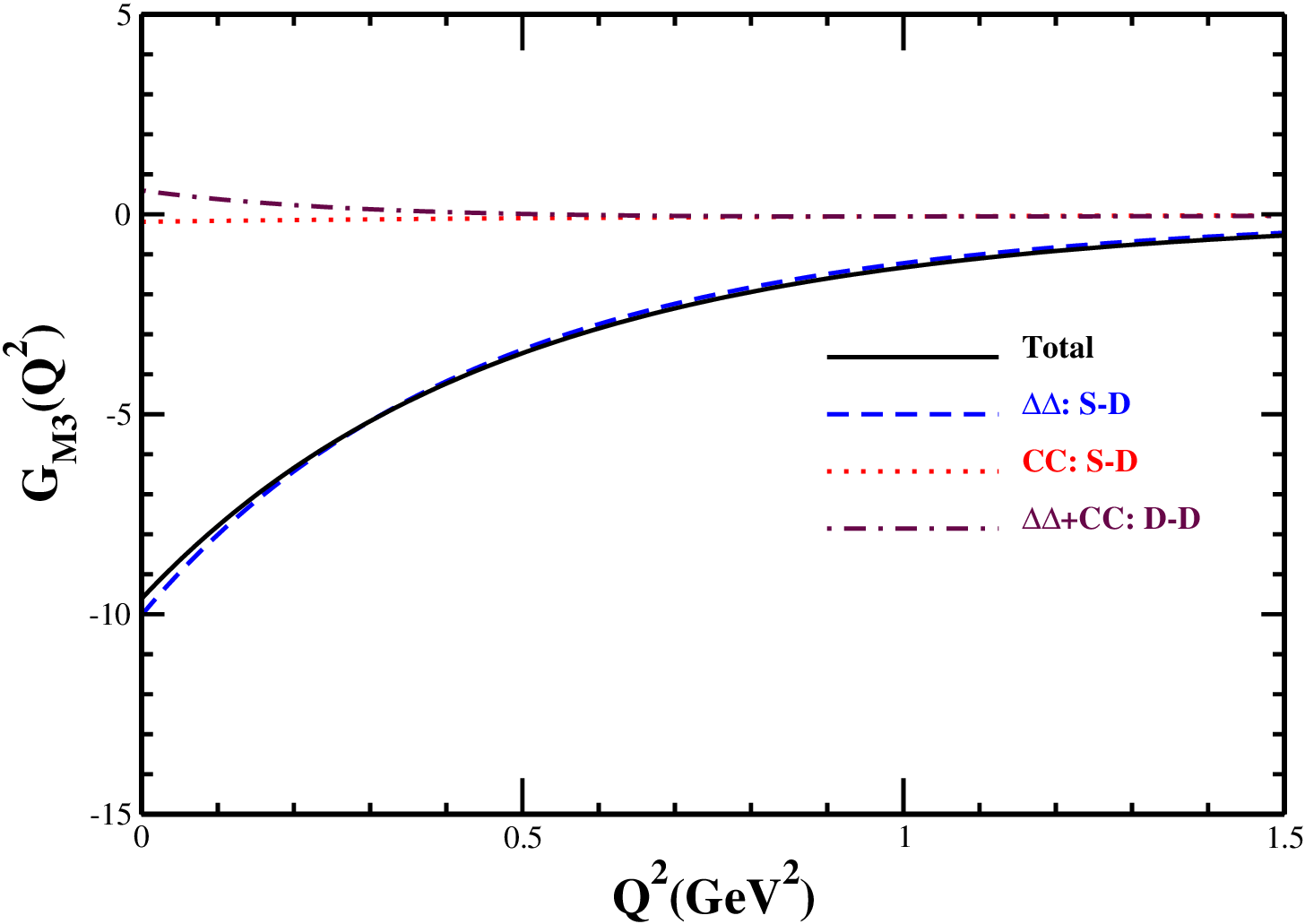}
\end{center}
\caption{\label{Fig4}The magnetic quadrupole form factors $M3$ of
$\ds$.}
\end{figure}
%\vspace{0.5cm}
Finally, we also shown the six-pole magnetic form factor of $d^*$  in Fig.~\ref{Fig4}, where the blue-dashed,
red-dotted, and purple-dotted-dashed curves represent the contributions from the matrix elements between the
$S$- and $D$-waves of the $\Delta\Delta$ and $CC$ components and between the $D$-waves of the whole
$\Delta\Delta+CC$ wave function. Still, the major contribution comes from the $S-D$ matrix element of
the $\Delta\Delta$ component as well. \\

\section{Summery}
\noindent\par

In order to understand the internal structure of the $\ds$ resonance discovered by CELSIUS/WASA and WASA@COSY
Collaborations, two major structural schemes were proposed recently. One of them considers that it has a compact
exotic hexaquark dominated structure and the other proposal believes that it is basically a molecular-like hadronic
state. These two structure models have been tested in terms of the experimental data. Up to now, both models can
explain the mass, the total width, and the partial decay widths for all the observed double pion decays of the
$\ds$ resonance. However, for a single pion decay process, although the observed upper limit of the branching ratio
can be explained by both structure models,  the ways of explanation have a subtle difference. The result from a
compact hexaquark dominated structure model is directly calculated and is consistent with the
data. On the other side, a combined $\alpha~[\Delta\Delta]+(1-\alpha)~[D_{12}\pi]$ mixing structure is proposed,
and the data can also be explained by fitting the value of $\alpha=5/7$ to the observed upper limit of the
branching ratio, because the result from a original $D_{12}\pi$ (or $\Delta N \pi$) structure model is excluded by
the data. Therefore, we need to seek other observable physical quantities to distinguish these two different
structures for $\ds$.\\

Here, based on the studies on the electromagnetic form factors for the nucleon, deuteron, and even vector mesons,
we propose that the electromagnetic form factors, including the $\ds$ charge distribution in our former paper
~\cite{Dong:2017mio}, can be the desirable physical quantities for distinguishing different structure
approaches. In this paper, we study the $M1$, $E2$, and $M3$ form factors in addition to the former reported
charge form factor $C0$ by employing the wave functions obtained in the coupled $\Delta\Delta+CC$ channel RGM
calculation based on our chiral constituent quark model. It is found that in the case with a compact
$\Delta\Delta+CC$ structure, since the D-wave components in both $\Delta\Delta$\ and $CC$ channels are negligible
small, less than $0.5$\% of the total wave function, its contribution to the electromagnetic form factor $M1$ is
rather small in comparison with that from the $S$-wave component. However, for the electromagnetic form factors
$E2$ and $M3$, the contribution of the $D$-wave associating with the $S$-wave, namely the off-diagonal matrix
elements between the $D$- and $S$-waves, of the $\Delta\Delta$ component dominates. The extracted magnetic
dipole moments of $\ds$ for the compact hexaquark dominated ($\Delta\Delta+CC$) structure, the pure $\Delta\Delta$
structure, and the $D_{12}\pi$ structure are $7.602$, $7.612$, and $~5.07$, respectively. The corresponding
magnetic radii are about $0.66~\rm{fm}$ in the case with a coupled $\Delta\Delta+CC$ structure and about
$0.90~\rm{fm}$ in the case with a single $\Delta\Delta$ structure, respectively. These results indicate that the
magnetic moment can be used to distinguish between the compact hexaquark dominated structure (or the pure
$\Delta\Delta$ structure) and the $D_{12}\pi$ structure, but not between the compact hexaquark dominated
structure and the pure $\Delta\Delta$ structure. However, the magnetic radius can be considered as a physical
quantity to discriminate the $\Delta\Delta+CC$ and $\Delta\Delta$ structures. Moreover, a quite small quadrupole
deformation $\hat{Q}_{20}^{\ds}$ of $2.53\times 10^{-2}~\rm{fm}^2$ for the $\ds$ state with a $\Delta\Delta+CC$
structure indicates that, differing with deuteron, $\ds$ is more inclined to a slightly oblate shape, and
consequently, a compact hexaquark dominated and spherical structure. The effect of the deformation of
the $\Delta$ resonance provides a suppression of about 0.25\% to the quadrupole moment of $\ds$.\\

Combining the results for $C0$ in our previous
paper~\cite{Dong:2017mio}, we come to the conclusion that the charge
radius and magnetic moment of $\ds$ can be used as new physical
quantities to discriminate among different structure models. It is
expected that these theoretically predicted quantities, especially
in the low-$Q$ region, can be measured by experiments in the near
future. For instance~\cite{Bashkanov:2018}, at Belle II, with its
high luminosity, it might be possible to access $e^++e^-\to
d^*+\bar{d}^*$, and then one might extract some information about the
electromagnetic form factors of $\ds$ in the time-like region.
Considering the photo-production process, one might access magnetic
moment information by photo-exciting the $P$-shell nucleon pair with
the M1 transition as well. There is another possible chance to
directly access the information of the electromagnetic feature of
$\ds$ at the low-$Q$ region, where one may look for a $e^+e^-$ pair
production process ($pn\to \ds e^+e^-$) at the WASA-at-GSI and CBM
due to the advantages of their deuteron beam
and a very good di-lepton efficiency and triggering in the CBM experiment.\\

\section*{Acknowledgment}

\noindent\par This work is supported by the National Natural
Sciences Foundations of China under the grant Nos. 11475192,
11475181, 11521505, 11565007, and 11635009, the Sino-German CRC 110
"Symmetries and the Emergence of Structure in QCD" project by NSFC
under the grant No.11621131001, the Key Research Program of Frontier
Sciences, CAS, Grant No. Y7292610K1, and the IHEP Innovation Fund
under the grant No. Y4545190Y2. Authors thank the fruitful
discussions with Mikhail Bashkanov and Yubing Dong thanks Fei Huang for
providing the wave functions of $\ds$.

\end{document}